\documentclass[conference]{IEEEtran}

\usepackage[utf8]{inputenc}
\usepackage[T1]{fontenc}

\usepackage{algorithm2e}
\usepackage{amsmath}
\usepackage{amssymb}
\usepackage{booktabs}
\usepackage{fancyvrb}
\usepackage{tabularx}

\newtheorem{lemma}{Lemma}
\newtheorem{theorem}{Theorem}
\newtheorem{example}{Example}

\makeatletter
\let\old@algorithm=\algorithm
\let\old@endalgorithm=\endalgorithm
\renewenvironment{algorithm}{\let\@latex@error\@gobble\old@algorithm[H]}{\old@endalgorithm}
\makeatother

\title{Determining the Multiplicative Complexity of Boolean Functions using SAT}
\author{%
  \IEEEauthorblockN{Mathias Soeken}
  \IEEEauthorblockA{Microsoft, Switzerland}
}

\begin{document}

\maketitle

\begin{abstract}
We present a constructive SAT-based algorithm to determine the multiplicative
complexity of a Boolean function, i.e., the smallest number of AND gates in any
logic network that consists of 2-input AND gates, 2-input XOR gates, and
inverters. In order to speed-up solving time, we make use of several symmetry
breaking constraints; these exploit properties of XAGs that may be useful
beyond the proposed SAT-based algorithm. We further propose a heuristic
post-optimization algorithm to reduce the number of XOR gates once the optimum
number of AND gates has been obtained, which also makes use of SAT solvers. Our
algorithm is capable to find all optimum XAGs for representatives of all
5-input affine-equivalent classes, and for a set of frequently occurring
6-input functions.
\end{abstract}

\section{Introduction}
We are considering the minimization of AND gates in XOR-AND graphs (XAGs), which
are logic networks that can have 2-input AND gates and 2-input XOR gates.  XAGs
can represent all normal Boolean functions, also called 0-preserving functions,
which are all Boolean functions $f$ for which $f(0, \dots, 0) = 0$.  A
non-normal Boolean function $f(x)$ can be represented by an XAG for $\bar f(x)$
with its output being inverted.  Note that the use of inverters cannot lead to a
smaller number of AND gates in the logic network~\cite{Schnorr88}.  Inverters
can always be propagated towards the outputs, since
\begin{multline}
  \bar x \oplus \bar y = x \oplus y,\;
  \bar x \oplus y = \overline{x \oplus y}, \\
  \bar xy = xy \oplus y,\; \text{and}\;
  \bar x\bar y = \overline{xy \oplus x \oplus y}.
\end{multline}
Missing cases are covered by commutativity of the operations.  In the remainder
of this paper, we assume that all Boolean functions are normal, unless
explicitly stated otherwise.

The multiplicative complexity of a Boolean function is the smallest number of
AND gates in any XAG that represents the function~\cite{Schnorr88}.  Determining
the multiplicative complexity is intractable.  Find has shown that if one-way
functions exist~\cite{Levin03}, no algorithm can compute the multiplicative
complexity of an $n$-variable Boolean function in time $2^{O(n)}$ given as input
the truth table of $f$~\cite{Find14}.  Therefore, heuristic optimization
techniques that minimize the number of AND gates in XAGs have been used as a
tool to assess the multiplicative complexity of the
function~\cite{BMP13,TSAM19,RJHK19,CCDE19,TSR+20}, since it provides an upper
bound of the actual complexity.

The concept of multiplicative complexity and the optimization of AND gates in
XAGs play an important role in cryptography and fault-tolerant quantum
computing.  In cryptography, the number of AND gates correlates to the degree of
vulnerability of a circuit~\cite{TP14}.  Further, the multiplicative complexity
of a function directly correlates to the resistance of the function against
algebraic attacks~\cite{CHM11}.  The number of AND gates also plays an important
role in high-level cryptography protocols such as zero-knowledge protocols,
fully homomorphic encryption (FHE), and secure multi-party computation
(MPC)~\cite{ARS+15,CDG+17,RJHK19}. For example, the size of the signature in
post-quantum zero-knowledge signatures based on ``MPC-in-the-head''~\cite{GMO16}
depends on the multiplicative complexity in the underlying block
cipher~\cite{CDG+17}. Moreover, the number of computations in MPC protocols
based on Yao’s garbled circuits~\cite{SHS+15} with the free XOR
technique~\cite{KS08} is proportional to the number of AND gates. Regarding FHE,
XOR gates are considered cheaper and less noisy compared to AND gates. In
fault-tolerant quantum computing, the multiplicative complexity directly
corresponds to the number of qubits and expensive operations ($T$ gates), and
therefore optimizing AND gates in an XAG can lead to more efficient quantum
circuits~\cite{MSC+19}.

In this paper, we present a SAT-based algorithm to determine the multiplicative
complexity of a Boolean function. The algorithm is constructive and returns an
XAG, in which the number of AND gates is minimum. The algorithm can only be
applied to either small functions or to functions which have a small
multiplicative complexity. It is inspired by SAT-based exact logic synthesis
techniques~\cite{KKY09,Knuth4f6,HSMM00}. While XAGs with minimum number of AND
gates are known for all 6-input functions~\cite{CTP19}, our algorithms can also
be used to enumerate multiple structurally different solutions, in order to
minimize the number of XOR gates, while keeping the number of AND gates
unchanged. For this purpose, we exploit an existing SAT-based algorithm to
minimize the number of XOR gates in logic networks for linear
functions~\cite{FS10}. Our algorithm is capable to find all optimum XAGs for
representatives of all 5-input affine-equivalent classes, and for a set of
frequently occurring 6-input functions. We further used the algorithm to find a
new collection of optimum XAGs for 5-input functions, categorized with respect
to representatives of a recently proposed affine equivalence classification
algorithm~\cite{STM19}.

\section{Preliminaries}
As general notation, we are using $[n] = \{1, \dots, n\}$. We model an XAG for a
Boolean function over variables $x_1, \dots, x_n$ as a sequence of $r$ steps,
where each step has one of the two following forms:
\begin{equation}
  x_i = x_{j_{1i}} \oplus x_{j_{2i}}
  \quad\text{or}\quad
  x_i = x_{j_{1i}} \land x_{j_{2i}}
\end{equation}
for $n < i \le n + r$.  The values $1 \le j_{1i} < j_{2i} < i$ point to primary
inputs or previous steps in the network.  The function value is computed by the
last step $f = x_{n+r}$.  We assume that all logic functions represent Boolean
functions that depend on all primary input variables.  To represent the constant
function, we define $x_0 = 0$.
\begin{example}
  \label{ex:ite-1}
  The \emph{if-then-else} function $x_1 \mathbin{?} x_2 \mathbin{:} x_3 = x_1x_2
  \oplus \bar x_1x_3 = x_1x_2 \oplus x_1x_3 \oplus x_3$ can be computed by the
  4-step XAG
  \[
    x_4 = x_1 \land x_2, \;
    x_5 = x_1 \land x_3, \;
    x_6 = x_4 \oplus x_5, \;
    x_7 = x_3 \oplus x_6.
  \]
\end{example}
The \emph{multiplicative complexity of an XAG} is the number of AND gates it
contains.  It is an upper bound to the multiplicative complexity of the Boolean
function, which is the smallest number of AND gates in any XAG that realizes the
function.
\begin{example}
  The multiplicative complexity of the XAG to realize the if-then-else function
  in Example~\ref{ex:ite-1} is 2, however, the function has multiplicative
  complexity 1, witnessed by the following alternative XAG:
  \[
    x_4 = x_2 \oplus x_3, \;
    x_5 = x_1 \land x_4, \;
    x_6 = x_3 \oplus x_5
  \]
\end{example}

Let $S = \{i_1, \dots, i_k\} \subseteq [n]$, then
\begin{equation}
  L_S(x_1, \dots, x_n) = x_{i_1} \oplus \cdots \oplus x_{i_k}
\end{equation}
is a linear function over the variables indexed by $S$.  We define
$L_{\emptyset} = 0$, and omit brackets when explicitly writing the indexes,
e.g., we write $L_{1,3}(x_1, x_2, x_3)$ instead of $L_{\{1,3\}}(x_1, x_2, x_3)$.

Since we are only interested in the number of AND gates in the XAG, we can
describe the networks in a more general way such that the number of steps equals
the number of AND gates.  Each fanin of an AND gate is a multi-input XOR gate
whose fanins in turn are either primary inputs or AND gates defined in previous
steps.  These multi-input XOR gates may only have a single input, in case the
fanin of an AND gate directly connects to a primary input or another AND gate.
Therefore, for a function $f(x_1, \dots, x_n)$ a more abstract logic network,
which we call abstract XAG in the following, consists of steps
\begin{equation}
  x_i = L_{S_{1i}} \land L_{S_{2i}}
\end{equation}
for $n < i \le n+r$ with $S_{1i}, S_{2i} \subseteq [i-1]$. The function value is
computed as a linear function over all primary inputs and AND gates $f = L_S$,
with $S \subseteq [n+r]$.  This abstract logic network has been defined in a
similar way by the authors in~\cite{CTP19}.  In order to simplify some notation,
we combine all index sets for linear functions by sets $\hat S_l$, where $\hat
S_1 = S_{1(n+1)}, \hat S_2 = S_{2(n+1)}, \hat S_3 = S_{1(n+2)}, \hat S_4 =
S_{2(n+2)}, \dots, \hat S_{2r+1} = S$.

\begin{example}
  The two XAGs for the if-then-else function from the previous examples can
  be represented by the abstract XAGs
  \[
    x_4 = L_1 \land L_2, \;
    x_5 = L_1 \land L_3, \;
    f = L_{3,4,5},
  \]
  and
  \[
    x_4 = L_1 \land L_{2,3}, \;
    f = L_{3,5},
  \]
  respectively.
\end{example}

It is straightforward to translate an XAG into an abstract XAG by merging all the
XOR gates between AND gates into single linear functions.  The inverse
direction, however, is more complicated.  A na\"ive translation of an abstract
XAG into an XAG may lead to a large number of XOR gates, if gate sharing and
cancellations are not taken into account.

\section{Finding Optimum Abstract XAGs}
We solve the optimization problem of finding the multiplicative complexity of a
normal Boolean function $f$, by solving a series of decision problems that ask
whether there exists an abstract XAG for $f$ with $r$ gates.  In that series of
decision problems, we can either increment $r$, starting from some lower bound,
or decrementing $r$, starting from an upper bound.  A good lower bound is $d -
1$, where $d$ is the algebraic degree of $f$, which is the size of the largest
monomial in its algebraic normal form~\cite{Schnorr88}.  A good upper bound is
the number of AND-like gates in a logic network for $f$ over binary gates.

\subsection{SAT Encoding}
We encode the decision problem as a SAT problem over variables $s_{cij}$ for $c
\in \{1,2\}$, $n < i \le n + r$ and $1 \le j < i$ that are true, whenever $j \in
S_{ci}$, and similarly, variables $s_j$ that are true, whenever $j \in S$ for
all $1 \le j \le n + r$. Further variables $f_{ix}$ for $n < i \le n+r$ and $0 <
x < 2^n$ encode the function computed by AND gate $i$ for input assignment $b_1,
\dots, b_n$ when $x = (b_n \dots b_1)_2$.  It is not necessary to consider the
case $x = 0$, since $f$ is normal. As notation we also introduce variables $\hat
s_{lj}$ which correspond to variables $s_{cij}$ and $s_j$ with respect to the
definition of the sets $\hat S_l$ above.  Note that these are not additional
variables for the SAT problem.

Clauses are added to relate the variables that encode the structure of the
abstract XAG with the variables that encode the function of the abstract XAG.
For each $x = (b_n \dots b_1)_2$, we add the gate clauses
\begin{equation}
  \label{eq:main-step}
  f_{ix} \leftrightarrow \bigwedge_{c\in\{1,2\}}
  \bigoplus_{j=1}^{i-1}(s_{cij} \land f_{jx}),
\end{equation}
for all $n < i \le n + r$.  Here, $f_{jx} = b_j$ whenever $j \le n$.  We also
add output clauses
\begin{equation}
  \label{eq:main-output}
  f(b_1, \dots, b_n) \leftrightarrow
  \bigoplus_{j=1}^{n+r}(s_{j} \land f_{jx}).
\end{equation}
Note that $f(b_1, \dots, b_n)$ evaluates to a constant value. All clauses are
added to the SAT solver by making use of the Tseytin encoding~\cite{Tseytin70}.

\subsection{Additional constraints and symmetry breaking}
The clauses in the previous section are sufficient to find abstract XAGs with a
maximum number of AND gates $r$ for some given function $f$.  In this section,
we introduce additional constraints that help the SAT solver, e.g., by using
redundant clauses, or by filtering out some solutions using symmetry breaking.

\subsubsection*{Non-constant linear fan-in}
Unless $f$ is the constant-0 function, all linear functions involved in the
abstract XAG should contain at least one variable, i.e., $\hat S_l \neq
\emptyset$ for all $l$, since they would otherwise evaluate to the $0$ function.
We can help the SAT solver by adding the clauses
\begin{equation}
  (\hat s_{l1} \lor \cdots \lor \hat s_{l|S_l|})
\end{equation}
for all $1 \le l \le 2r + 1$.

\subsubsection*{Commutativity}
Since the AND operation is commutative, we can interchange $S_{1i}$ and $S_{2i}$
in any step without changing the function.  We can force the SAT solver to break
this symmetry by asserting that $S_{1i}$ must be lexicographically smaller than
$S_{2i}$, i.e., $\max \{S_{1i} \triangle S_{2i} \} \in S_{2i}$, where
`$\triangle$' denotes the symmetric difference.  This constraint can be enforced
by adding the clauses

\begin{footnotesize}
\begin{multline*}
\bigwedge_{j=1}^{i-2} \big(
(\bar s_{1ij} \lor s_{2ij} \lor \bar a_{i(j-1)})
(\bar s_{1ij} \lor a_{ij} \lor \bar a_{i(j-1)})
(s_{2ij} \lor a_{ij} \lor \bar a_{i(j-1)})
\big) \\
(\bar s_{1i(i-1)} \lor \bar a_{i(i-2)})
(s_{2i(i-1)} \lor \bar a_{i(i-2)}),
\end{multline*}
\end{footnotesize}

\noindent for $n < i \le n+r$, and $i-2$ auxiliary variables $a_{ij}$, where all
literals $\bar a_{i0}$ are omitted~\cite{Knuth4f6}.

\subsubsection*{Symmetric variables}
A function $f(x_1, \dots, x_n)$ is symmetric in two variables $x_j$ and $x_k$,
if swapping $x_j$ with $x_k$ in $f$ does not change the function.  In that case,
we can enforce that $x_j$ ``is used before'' $x_k$, by adding the clauses
\[
  \hat s_{lk} \rightarrow \bigvee_{1 \le l' \le l} \hat s_{l'j}
\]
for $1 \le l \le 2r + 1$.

\subsubsection*{All gates and all essential variables must be used}
A variable $x_i$ is essential in $f$ if $f_{\bar x_i} \neq f_{x_i}$, where
$f_{\bar x_i}$ and $f_{x_i}$ are the negative and positive co-factors of $f$,
which are obtained by setting $x_i$ to $0$ and $1$, respectively.  In an optimum
solution each essential variable and each AND gate must be present in at least
one of the index sets $\hat S_l$.  This can be enforced by the clauses
\begin{equation}
  \bigvee_{l \text{ s.t. } |\hat S_l| \ge j} \hat s_{li}
\end{equation}
for all $i \in [n]$ such that $x_i$ is an essential variable, and for all $n < i
\le n + r$.

\subsubsection*{Linear fan-ins are no subsets}
It can be shown that there always exists a minimum abstract XAG in which $S_{1i}
\not\subseteq S_{2i}$ and $S_{2i} \not\subseteq S_{1i}$ for all steps $i$.  This
property has also been used by the authors in~\cite{CTP19} to reduce the number
of topologies in the enumeration of minimum abstract XAGs for all 6-variable
Boolean functions. The proof makes use of the following lemma:
\begin{lemma}
  \label{lemma:subset-s}
  Let $S_1 \subseteq S_2$, then
  \begin{equation}
    L_{S_1} \land L_{S_2} = L_{S_1} \oplus (L_{S_1} \land L_{S_2 \setminus S_1})
  \end{equation}
\end{lemma}
\begin{IEEEproof}
  We start by expanding the left-hand side of the equality using the
  distributivity law:
  \[
    L_{S_1} \land L_{S_2} = \bigoplus_{(i,j) \in S_1\times S_2} x_ix_j
  \]
  Splitting $S_2$ into the disjoint union $S_1 \cup (S_2 \setminus S_1)$ leads
  to
  \[
    \bigoplus_{(i,j) \in S_1 \times S_1} x_ix_j \oplus
    \bigoplus_{(i,j)\in S_1\times (S_2\setminus S_1)} x_ix_j.
  \]
  Since for each pair $(i, j) \in S_1 \times S_1$ with $i \neq j$, there also
  exists $(j, i) \in S_1 \times S_1$, all monomials of degree~2 on the left-hand
  side cancel, simplifying the expression to
  \[
    \bigoplus_{i \in S_1} x_i \oplus
    \bigoplus_{(i,j)\in S_1\times (S_2\setminus S_1)} x_ix_j.
  \]
  Now, it is easy to see that the term on the left-hand side is the linear
  function $L_{S_1}$, and the term on the right-hand side equals $L_{S_1} \land
  L_{S_2\setminus S_1}$ by laws of distributivity.
\end{IEEEproof}

We can now formulate and prove the following theorem.
\begin{theorem}
  \label{theo:subset-s}
  Any abstract XAG can be expressed using steps in which $S_{1i} \not\subseteq
  S_{2i}$ and $S_{2i} \not\subseteq S_{1i}$ without changing the number of
  steps.
\end{theorem}
\begin{IEEEproof}
  Assume we have an abstract XAG with $r$ steps that does not fulfill the stated
  property.  That is, there exists some step $i$ for which $S_{1i} \subseteq
  S_{2i}$ or $S_{2i} \subseteq S_{1i}$.  Without loss of generality, let us
  assume that $S_{1i} \subseteq S_{2i}$.  Then due to
  Lemma~\ref{lemma:subset-s}, we can rewrite that step as
  \[ x_i = L_{S_{1i}} \land S_{S_{2i} \setminus S_{1i}}, \] and replace all $S'
  \in \{S\} \cup \{S_{1j}, S_{2j} \mid j > i \}$ by $S' \triangle S_{1i}$,
  whenever $i \in S'$.  We repeat this process for all gates that do not fulfill
  the property, and since a change only affects gates with a larger index, this
  procedure eventually terminates.
\end{IEEEproof}

Based on this result, we add the clauses
\begin{equation}
  \bigvee_{j=1}^{i-1}(s_{1ij} \land \bar s_{2ij})
  \land
  \bigvee_{j=1}^{i-1}(\bar s_{1ij} \land s_{2ij})
\end{equation}
for all $n < i \le n+r$ to the SAT formula, thereby ruling out abstract XAGs
with the property in Theorem~\ref{theo:subset-s}.

\subsubsection*{Multi-level subset relation}
More symmetry breaking can be taken into account when considering subset
relations among linear functions that are not from the same AND gate.  For this
purpose, we make use of the following lemma, which is proven in
Appendix~\ref{ap:lemma-subsets}.
\begin{lemma}
  \label{lemma:lemma-subsets}
  Let $S$, $T$, and $U$ such that $S \subseteq T$, and $S \subseteq U$.  Then
  \begin{equation}
    \label{eq:lemma-subsets}
    L_S \oplus (L_T \land L_U) =
    L_{T\setminus S} \oplus (L_T \land L_{T\triangle U}).
  \end{equation}
\end{lemma}
On the left-hand side, by definition $S \subseteq T$ and $S \subseteq U$.
However, on the right-hand side, $T\setminus S \subseteq T$, but $T\setminus S
\subseteq T \triangle U$, if and only if $T \cap U = S$.  Therefore, we can
restrict the number of solutions, by constraining that $S \subseteq T$ and $S
\subseteq U$ implies that $T \cap U = S$, by adding the clauses
\begin{multline}
  \bigwedge_{i=n+1}^{|\hat S_l|}
  \Bigg(\Big(
    \hat s_{li} \land
    \bigwedge_{j=1}^{i-1}(\hat s_{lj} \rightarrow s_{1ij}s_{2ij})
  \Big) \\
  \rightarrow
  \bigwedge_{j=1}^{i-1}(\hat s_{lj} \leftrightarrow s_{1ij}s_{2ij})
  \Bigg)
\end{multline}
for all $1 \le l \le 2r+1$. In that case, the SAT solver, will rule out the
left-hand side expression in~\eqref{eq:lemma-subsets}, unless $T \cap U = S$.

\subsection{Solving strategies}
\label{sec:strategies}
In this section we describe different solving strategies to solve the
optimization problem of finding an XAG with the minimum number of AND gates for
an $n$-variable Boolean function $f$.  For this purpose, we make use of the
following notation. Function $F_r(x)$ describes the conjunction of the main
clauses~\eqref{eq:main-step} and~\eqref{eq:main-output} for some given number of
steps $r$ and input assignment $x$.  The function $A_r$ is a conjunction of
additional and symmetry breaking constraints described in the previous section.
Note that $A_r$ does not necessarily need to contain all of these constraints.
The value $r_{\mathrm{low}}$ is a lower bound for $r$, and is either set to $d -
1$, where $d$ is the algebraic degree of $f$, or an application specific value.
The function $\mathrm{solve}$ solves a SAT formula using a SAT solver and
returns either $\mathrm{sat}$ or $\mathrm{unsat}$, and the function
$\mathrm{extract\_xag}$ extracts an XAG from the last satisfying SAT call.

\subsubsection*{Direct method} The direct method solves the optimization problem
as follows:

\begin{algorithm}
set $r\gets r_{\mathrm{low}}$\;
\While{$\mathrm{solve}(\bigwedge_{x=1}^{2^n-1}F_r(x) \land A_r) = \mathrm{unsat}$}{
  set $r\gets r + 1$\;
}
\Return{$\mathrm{extract\_xag}()$}\;
\end{algorithm}
Starting from a lower bound, it constrains all input assignments (except $x =
0$), and returns one XAG for the first satisfying solution.

\subsubsection*{Counter-example guided abstraction refinement}
In this method inspired by~\cite{CGJLV03} we do not constrain all input
assignments, but call the SAT solver incrementally, constraining new input
assignments that are derived from counter-examples of wrong solutions:

\begin{algorithm}
set $r \gets r_{\mathrm{low}}$\;
\While{true}{%
  set $F \gets A_r$\;
  \While{$\mathrm{solve}(F) = \mathrm{sat}$}{
    set $N \gets \mathrm{extract\_xag()}$\;
    set $f' \gets \mathrm{simulate}(N)$\;
    \uIf{$f' = f$}{
      \Return{$N$}\;
    }
    \Else{
      set $x \gets \min_x \{ f(x) \neq f'(x) \}$\;
      set $F \gets F \land F_r(x)$\;
    }
  }
  set $r \gets r + 1$\;
}
\end{algorithm}

After each satisfying SAT call, we extract a candidate XAG $N$ and simulate it
to extract its function $f'$.  If $f' = f$, then $N$ is a minimum XAG, otherwise
we refine $F$ by constraining the smallest input assignment in which $f$ and
$f'$ differ.

\subsubsection*{Multiple solutions}  Next, we assume that we have found some
$r$, for which there exists an XAG that implements $f$.

\begin{algorithm}
  set $\mathcal{N} \gets \emptyset$\;
  set $F \gets \bigwedge_{x=1}^{2^n-1}F_r(x) \land A_r$\;
  \While{$\mathrm{solve}(F) = \mathrm{sat}$}{
    set $\mathcal{N} \gets \mathcal{N} \cup \{\mathrm{extract\_xag()}\}$\;
    set $F \gets F \land \mathrm{block()}$\;
  }
  \Return{$\mathcal{N}$}\;
\end{algorithm}

While $F$ is satisfiable and solutions can be extracted, we iteratively block
solutions, by adding a single large clause that contains all variables $\hat
s_{lj}$ in a polarity opposite to their value in the last satisfying solution.

\section{Optimizing the Number of XOR Gates}
There exist several differently structured abstract XAGs, all having the same
number of steps, and all realizing the same Boolean function $f$.  Further, for
each of these abstract XAGs there exists several possible ways to transform it
into an XAG without changing its structure, i.e., without changing the linear
functions that fan-in to the AND gates. Finding the XAG with the smallest number
of XOR gates for a given number of AND gates essentially requires to solve the
SAT problem on the XAG structure rather than on the abstract XAG, which makes
the problem much more complicated.  Instead, in this section, we propose a
method that (i) heuristically determines an abstract XAG that potentially leads
to a small number of XOR gates in the XAG, and (ii) finds an XAG with the
smallest number of XOR gates for that particular abstract XAG exactly using a
SAT-based method.

\subsection{Heuristic to find good abstract XAG as starting point}
\label{sec:xor-heuristic}
The number of XOR gates is possibly small, if the linear functions in the
abstract XAG have small index sets.  In other words, we aim to minimize the sum
$|\hat S_1| + \cdots + |\hat S_{2r + 1}|$.  For this purpose, we add the
constraint
\begin{equation}
  \label{eq:card-p}
  \sum_{l=1}^{2r+1}\sum_{j=1}^{|\hat S_l|} \hat s_{lj} < p
\end{equation}
to an otherwise satisfying instance to find an abstract XAG with $r$ gates,
where $r$ is already known to be optimum.  An initial value for $p$ can be
extracted from a known solution, and then incrementally made smaller until no
more satisfying solution can be found.  It is a common approach to implement
such a search using a sorter network, which has as inputs all variables $\hat
s_{lj}$.  We can then constrain~\eqref{eq:card-p} by forcing the
$p^{\mathrm{th}}$ most significant output of the sorter network to be
$0$~\cite{CZ10}.

\subsection{Finding the shortest linear network with SAT}
\label{sec:xor-sat}
Next we describe a SAT-based approach that finds the smallest XAG representation
in terms of XOR gates for a given abstract XAG.  For this purpose, we introduce
multi-ouput linear networks.  Let $L_{S_1}, \dots, L_{S_m}$ be $m$ linear
functions over $n$ variables.  We can represent all functions by a single $m
\times n$ Boolean matrix $A = (a_{lj})$ with $a_{lj} = [j \in S_l]$.  A linear
network (or linear straight-line program) for $m$ linear functions over $n$
variables is a sequence of steps
\begin{equation}
  x_i = x_{j_{1i}} \oplus x_{j_{2i}}
\end{equation}
for $n < i \le n + r$ and a mapping $f_l \in [n+r]$ for $1 \le l \le m$.

Given an abstract XAG over $n$ variables and $r$ steps, we can extract a linear
network over $n+r$ variables and $2r + 1$ linear functions, where variables
$x_1, \dots, x_{n+r}$ correspond to the original inputs and steps in the
abstract XAG and the $2r + 1$ linear functions are $S_{1(n+1)}, S_{2(n+1)},
\dots, S_{1(n+r)}, S_{2(n+r)}, S$, in that order.  From the linear network we
can then construct an XAG, where the inputs $x_{n+1}, \dots, x_{n+r}$ correspond
to the outputs of an AND gate, and where all but the last linear function
correspond to the inputs of an AND gate.  To ensure that the XAG corresponds to
a directed acyclic graph, we must control how steps are computed in the linear
network.  For example, in order to compute a function that corresponds to an
input of AND gate $x_i$, we cannot use any variable with index $i$ or higher.

For a single linear function $L_S$, where $S \neq \emptyset$, the shortest
linear network requires $r = |S| - 1$ steps. Determining the smallest linear
network for a set of linear functions is more complicated, since one needs to
take step sharing and step cancellations into account.  Boyar, Matthews, and
Peralta have shown that finding a linear network with the smallest number of
steps for some set of linear functions is MaxSNP-complete~\cite{BMP08}.

Fuhs and Schneider-Kamp have presented a SAT-based approach to find the shortest
linear network for a given Boolean matrix $A$~\cite{FS10}.  Similarly to the
approach presented in this paper, they are solving a sequence of decision
problems that ask whether there exists a linear network with $r$ steps.  Their
SAT encoding for this decision problem consists of variables $b_{ij}$ that
encode whether step $x_i$ uses input or step $x_j$ for $n < i \le n + r$ and $1
\le j < i$, and variables $f_{li}$ that encode whether step $x_i$ computes
output $f_l$ for $n < i \le n + r$ and $1 \le l \le m$.  For simplicity, we
assume that each row in $A$ has at least two 1s, i.e., no output is a constant-0
or projection function.  Then, the constraints
\begin{equation}
  \sum_{j=1}^{i-1} b_{ij} = 2 \quad\text{for $n < i \le n+r$}
\end{equation}
ensure that each step has two inputs and the constraints
\begin{equation}
  \sum_{i=n+1}^{n+r} f_{li} = 1 \quad\text{for $1 \le l \le m$}
\end{equation}
ensure that each output is computed by exactly one step.  The main clauses
\begin{equation}
  \bigwedge_{i=n+1}^{n+r}\left(
    f_{li} \rightarrow
    \bigwedge_{j=1}^n
    (\psi(j,i) \leftrightarrow a_{lj})
  \right)
  \quad\text{for $1 \le j \le m$}
\end{equation}
ensure that if the function $f_l$ is computed by step $x_i$, the linear function
at that step matches the entries in row $l$ of $A$.  For this purpose, the
recursive function
\begin{equation}
  \label{eq:psi}
  \psi(j,i) = b_{ij} \oplus \bigoplus_{i'=n+1}^{i-1} b_{ii'} \land \psi(j, i')
\end{equation}
encodes whether variable $x_j$ is in the linear function computed by step $x_i$
for $1 \le j \le n$ and $n < i \le n+r$.  Further constraints to remove
redundant solutions or break symmetries can speed up the SAT solver and are
discussed in~\cite{FS10}.

As described above, a linear network extracted from an abstract XAG must obey
additional constraints to ensure that linear functions that correspond to input
of gate $x_i$ do not make use of variables that correspond to the output of a
gate $x_j$ with $j>i$.  One possibility to ensure these constraints is to force
the linear network to be cancellation-free~\cite{BMP08}, however, this is
unnecessarily restrictive. A better way is to forbid using some inputs in the
computation of an output. For this purpose, we extend the original SAT encoding
to find the shortest linear network by the recursive function
\begin{equation}
  \varphi(j, i) = b_{ij} \lor \bigvee_{i' = n+1}^{i-1} b_{ii'} \land \varphi(j, i'),
\end{equation}
for $1 \le j \le n$ and $n < i \le n+r$, which is similar to $\psi(j, i)$
in~\eqref{eq:psi}, but captures whether variable $x_j$ has been used in any
intermediate step to compute $x_i$.  If we wish to enforce cancellation-free
logic networks, we can add the constraints $\psi(j, i) \leftrightarrow
\varphi(j, i)$, however, for our application it is sufficient to simply forbid
that some outputs cannot use some inputs by constraining
\begin{equation}
  f_{li} \rightarrow \bar b_{ij}
\end{equation}
for all $n < i \le n + r$, $1 \le l < m$ and $n + \lceil\frac{l}{2}\rceil \le j
\le n+r$.

\section{Experimental Results} We have implemented the algorithm using the EPFL
logic synthesis libraries~\cite{SRHM18}. In our experiments, we used the exact
synthesis algorithms to verify known multiplicative complexities for Boolean
functions with up to 6 variables~\cite{TP14,CTP19}. More precisely, we found
optimum XAGs for affine-equivalent classes, since the multiplicative complexity
is invariant for all functions in a class. As representative function, we chose
the function determined by the classification algorithm in~\cite{STM19}. The
experiment \textit{affine}$(n)$ finds optimum XMGs for all affine equivalent
class representatives for $n$-variable Boolean functions, starting from a lower
bound based on the function's algebraic degree. There are $8$ and $48$ classes
for all $4$- and $5$-variable Boolean functions,
respectively~\cite{Harrison64}. The experiment \textit{practical6}$(k)$ finds
optimum XAGs for the $k$ most occurring $6$-variable Boolean functions found by
enumerating cuts in all arithmetic and random control benchmarks of the EPFL
logic synthesis benchmark suite~\cite{AGM15}. Appendix~\ref{ap:practical6}
shows how these functions are obtained using ABC~\cite{BM10}.

We report the total number of 2-input XOR gates, the total number of 2-input AND
gates, and the overall run-time in seconds.  We perform four runs for each
experiment: The first run finds a single optimum XAG; the second run uses the
solving strategy from Sect.~\ref{sec:strategies} to find up to 50 optimum XAGs
and then selects the one with the fewest number of XOR gates; the third run also
finds up to 50 optimum XAGs but uses the XOR minimization heuristic based on
sorter networks presented in Sect.~\ref{sec:xor-heuristic}; the fourth run is
like the third run, but additionally optimizes XOR gates in the final solution
using the SAT-based approach in Sect.~\ref{sec:xor-sat}.

We run all the experiments on a Microsoft Azure virtual machine, on a general
purpose Standard D2 v3 size configuration, running on a Intel Xeon Platinum
8171M 2.60GHz CPU with 8 GiB memory and Ubuntu 18.04. We use Z3~\cite{MB08} as
a SAT solver. We also tried ABC's~\cite{BM10} modified implementation of
MiniSAT~\cite{ES03}, Glucose~\cite{LS09}, and MapleSAT~\cite{LGPC16}, which
both performed slower in these experiments. We set conflict limits for the SAT
solvers used in abstract XAG optimization and linear network optimization of
50,000 and 500,000 conflicts, respectively. These conflicts are only applied
once the first optimum solution has been found.

\begin{table}[t]
  \def\tabcolsep{6pt}
  \caption{Experimental Results}
  \label{tbl:experiments}
  \begin{tabularx}{\linewidth}{Xccrrr}
    \toprule
    Experiment & mult. & XOR opt. & \#XOR & \#AND & Runtime \\
    \midrule
    \textit{all-affine}$(4)$    &              &       &  44  &  16 &    0.13 \\
    \textit{all-affine}$(4)$    & $\checkmark$ &       &   9  &  16 &    0.93 \\
    \textit{all-affine}$(4)$    & $\checkmark$ & heur. &   5  &  16 &    0.45 \\
    \textit{all-affine}$(4)$    & $\checkmark$ & SAT   &   5  &  16 &    0.48 \\[2pt]
    \textit{all-affine}$(5)$    &              &       & 508  & 162 &  125.98 \\
    \textit{all-affine}$(5)$    & $\checkmark$ &       & 245  & 162 &  251.26 \\
    \textit{all-affine}$(5)$    & $\checkmark$ & heur. & 179  & 162 &  412.26 \\
    \textit{all-affine}$(5)$    & $\checkmark$ & SAT   & 173  & 162 &  530.10 \\[2pt]
    \textit{practical6}$(100)$  &              &       & 1161 & 293 &  558.95 \\
    \textit{practical6}$(100)$  & $\checkmark$ &       & 1022 & 293 &  663.76 \\
    \textit{practical6}$(100)$  & $\checkmark$ & heur. &  981 & 293 & 3888.42 \\
    \textit{practical6}$(100)$  & $\checkmark$ & SAT   &  771 & 293 & 5881.60 \\[2pt]
    \bottomrule
  \end{tabularx}
\end{table}

We further used our exact abstract XAG algorithm together with the XOR
minimization strategy to find small XAGs for all 48 affine equivalence classes
for 5-variable Boolean functions, such that the class representative matches the
function that is returned by the classification algorithm in~\cite{STM19}.
Table~\ref{tbl:all5} lists all XAGs together with the truth table representation
of each function in hexadecimal notation and also lists the corresponding
multiplicative complexity.  Such a table can be used in AND minimization
techniques such as the rewriting algorithm presented in~\cite{TSAM19}.

\section{Conclusions}
We presented a SAT-based algorithm to determine the multiplicative complexity of
a Boolean function.  The algorithm is only applicable to small Boolean
functions.  The multiplicative complexity is known for all Boolean functions up
to 6 variables, however, the approaches that were used to determine these
numbers are based on the exhaustive enumeration of all 150,357 equivalence
classes. Instead, our approach can be used to find one or multiple solutions for
one particular Boolean function of interest.  Therefore, for functions with up
to 6 variables, it is of interest when the goal is to explore different
structures for the same function.  For functions with more than 6 variables, it
can find solutions, if the multiplicative complexity is low, or can prove that
no solution with a small number of AND gates can exist.

Future insight in the structure of XAGs can be exploited as additional symmetry
breaking rules in order to speed up solving time.  Similarly, advances in SAT
solvers have a positive impact on our proposed algorithm.

\subsubsection*{Acknowledgments}
We like to thank Thomas H\"aner, Robin Kothari, Ren{\'e} Peralta, Michael
Miller, Bruno Schmitt, and Eleonora Testa for valuable feedback in the
preparation of this manuscript.

\bibliographystyle{IEEEtran}
\bibliography{library}

\appendix
\subsection{Proof of Lemma~\ref{lemma:lemma-subsets}}
\label{ap:lemma-subsets}
We introduce a notation and identities for a subset of quadratic forms, which
will be useful for the proof.  Quadratic forms are algebraic normal forms, in
which all monomials have degree~2.  A subset of quadratic forms can be
constructed for two disjoint sets $S$ and $T$, i.e., $S \cap T = \emptyset$.  We
define
\begin{equation}
  Q_{ST} = L_S \land L_T = \bigoplus_{(i,j) \in S\times T} x_ix_j.
\end{equation}
Note that $Q_{ST}$ contains of $|S| \times |T|$ monomials of degree~2, since no
cancellation can take place.  Also note that $Q_{ST} = Q_{TS}$.  Let $S_1, S_2$,
and $T$ such that all three sets are pairwise disjoint.  Then
\begin{equation}
  \label{eq:merge-quadratic}
  Q_{S_1T} \oplus Q_{S_2T} = Q_{(S_1 \cup S_2)T}.
\end{equation}
Further, for any two sets $S$ and $T$, we have
\begin{equation}
  \label{eq:merge-xor}
  L_S \oplus L_T = L_{S\triangle T}
\end{equation}
and
\begin{equation}
  \label{eq:merge-and}
  L_S \land L_T =
  L_{S \cap T} \oplus
  Q_{(S\cap T)(S \triangle T)} \oplus
  Q_{(S\setminus T)(T \setminus S)}
\end{equation}

We prove one non-trivial lemma, based on these identities:
\begin{lemma}
  \label{lemma:rhs}
  Given two sets $T$ and $U$, we have $L_T \land L_{T \triangle U} = L_T \oplus
  (L_T \land L_U)$.
\end{lemma}
\begin{IEEEproof}
  Since $T \cap (T\triangle U) = T\setminus U$, $T \triangle (T \triangle U) =
  U$, $T \setminus (T \triangle U)  T \cap U$, and $(T\triangle U)\setminus T =
  U\setminus T$, by expanding the left-hand side of the lemma
  using~\eqref{eq:merge-and}, we get
  \[
    L_{T\setminus U} \oplus Q_{(T\setminus U)U} \oplus Q_{(T \cap U)(U \setminus T)}.
  \]
  We add twice the term $L_{T\cap U}$, which does not change the function due to
  cancellation, and also expand the first quadratic form on $U = (T \cap U) \cup
  (U \setminus T)$ using~\eqref{eq:merge-quadratic}:
  \begin{multline*}
    L_{T\setminus U} \oplus L_{T\cap U} \oplus L_{T\cap U} \\ \oplus Q_{(T\setminus U)(T \cap U)} \oplus Q_{(T\setminus U)(U\setminus T)} \oplus Q_{(T \cap U)(U \setminus T)}.
  \end{multline*}
  Applying~\eqref{eq:merge-xor} to the first two linear forms,
  and~\eqref{eq:merge-quadratic} to the first and third quadratic form gives
  \[
    L_T \oplus L_{T\cap U} \oplus Q_{(T\cap U)(T \triangle U)} \oplus Q_{(T\setminus U)(U\setminus T)},
  \]
  which simplifies to $L_T \oplus (L_T \land L_U)$ after
  applying~\eqref{eq:merge-and}.
\end{IEEEproof}

We can now prove Lemma~\ref{lemma:lemma-subsets}, and show that~$L_{T\setminus
S} \oplus (L_T \land L_{T\triangle U}) = L_S \oplus (L_T \land L_U)$, if $S
\subseteq T$ and $S \subseteq U$.
\begin{IEEEproof}
  \[
    L_{T\setminus S} \oplus (L_T \land L_{T\triangle U})
    \stackrel{\text{Lemma~\ref{lemma:rhs}}}{=}
    L_{T\setminus S} \oplus L_T \oplus (L_T \land L_U)
  \]
  Since $S \subseteq T$, we have $(T\setminus S)\triangle T = S$, and
  applying~\eqref{eq:merge-xor} yields $L_S \oplus (L_T \land L_U)$.
\end{IEEEproof}

\subsection{Functions in \textit{practical6}$(n,k)$}
\label{ap:practical6}
The following procedure generates the $k$ practical $6$-variable functions in
\textit{practical6}$(n,k)$ using ABC~\cite{BM10} and the arithmetic and random
control instances of the EPFL logic synthesis benchmarks~\cite{AGM15}.

\begin{small}
\begin{Verbatim}[commandchars=\\\{\},codes={\catcode`$=3\catcode`_=8}]
\$ abc
abc> arithmetic/adder.aig; cut -s
abc> arithmetic/bar.aig; cut -s
...
abc> random\_control/router.aig; cut -s
abc> random\_control/voter.aig; cut -s
abc> npnsave functions; quit
\$ cat functions | grep -e "6\$" | head -n$k$ \\
                | cut -d ' ' -f1
\end{Verbatim}
\end{small}

\newcommand{\hex}[1]{\ensuremath{{}^{\scriptscriptstyle\#}\mathtt{#1}}}
\begin{table*}[t]
  \caption{Optimum XAGs for all 48 affine equivalent class for 5-variable Boolean functions}
  \label{tbl:all5}
  \footnotesize
  \def\tabcolsep{2pt}
  \renewcommand{\arraystretch}{1.2}
  \begin{tabularx}{\linewidth}{rl>{\scriptsize}Xr}
    \toprule
    Class & Function & {\footnotesize Logic network} & MC \\
    \midrule
     0 & \hex{00000000} & $f = x_0$ & 0 \\
     1 & \hex{80000000} & $x_{6} = x_{1} \land x_{2}, x_{7} = x_{3} \land x_{4}, x_{8} = x_{6} \land x_{7}, x_{9} = x_{5} \land x_{8}$ & 4 \\
     2 & \hex{80008000} & $x_{5} = x_{1} \land x_{2}, x_{6} = x_{3} \land x_{4}, x_{7} = x_{5} \land x_{6}$ & 3 \\
     3 & \hex{00808080} & $x_{6} = x_{1} \land x_{2}, x_{7} = x_{3} \land x_{6}, x_{8} = x_{4} \land x_{7}, x_{9} = x_{5} \land x_{8}, x_{10} = x_{7} \oplus x_{9}$ & 4 \\
     4 & \hex{80808080} & $x_{4} = x_{1} \land x_{2}, x_{5} = x_{3} \land x_{4}$ & 2 \\
     5 & \hex{08888000} & $x_{6} = x_{1} \land x_{2}, x_{7} = x_{3} \land x_{4}, x_{8} = x_{5} \oplus x_{7}, x_{9} = x_{6} \land x_{8}$ & 3 \\
     6 & \hex{aa2a2a80} & $x_{6} = x_{2} \land x_{3}, x_{7} = x_{4} \oplus x_{6}, x_{8} = x_{1} \land x_{7}, x_{9} = x_{4} \land x_{8}, x_{10} = x_{1} \oplus x_{9}, x_{11} = x_{5} \land x_{10}, x_{12} = x_{8} \oplus x_{11}$ & 4 \\
     7 & \hex{88080808} & $x_{6} = x_{1} \land x_{2}, x_{7} = x_{3} \land x_{6}, x_{8} = x_{4} \land x_{7}, x_{9} = x_{5} \land x_{8}, x_{10} = x_{6} \oplus x_{9}, x_{11} = x_{7} \oplus x_{10}$ & 4 \\
     8 & \hex{2888a000} & $x_{6} = x_{2} \land x_{5}, x_{7} = x_{3} \land x_{4}, x_{8} = x_{6} \oplus x_{7}, x_{9} = x_{1} \land x_{8}$ & 3 \\
     9 & \hex{f7788000} & $x_{6} = x_{3} \land x_{4}, x_{7} = x_{1} \land x_{2}, x_{8} = x_{5} \oplus x_{6}, x_{9} = x_{3} \oplus x_{4}, x_{10} = x_{5} \oplus x_{9}, x_{11} = x_{7} \oplus x_{10}, x_{12} = x_{8} \land x_{11}, x_{13} = x_{5} \oplus x_{12}$ & 3 \\
    10 & \hex{a8202020} & $x_{6} = x_{4} \land x_{5}, x_{7} = x_{3} \oplus x_{6}, x_{8} = x_{2} \land x_{7}, x_{9} = x_{3} \oplus x_{8}, x_{10} = x_{1} \land x_{9}$ & 3 \\
    11 & \hex{08880888} & $x_{5} = x_{1} \land x_{2}, x_{6} = x_{3} \land x_{5}, x_{7} = x_{4} \land x_{6}, x_{8} = x_{5} \oplus x_{7}$ & 3 \\
    12 & \hex{bd686868} & $x_{6} = x_{4} \land x_{5}, x_{7} = x_{2} \oplus x_{6}, x_{8} = x_{1} \land x_{7}, x_{9} = x_{2} \land x_{3}, x_{10} = x_{3} \oplus x_{8}, x_{11} = x_{1} \oplus x_{9}, x_{12} = x_{10} \land x_{11}, x_{13} = x_{6} \oplus x_{12}$ & 4 \\
    13 & \hex{aa808080} & $x_{6} = x_{2} \land x_{3}, x_{7} = x_{4} \land x_{5}, x_{8} = x_{6} \land x_{7}, x_{9} = x_{6} \oplus x_{7}, x_{10} = x_{8} \oplus x_{9}, x_{11} = x_{1} \land x_{10}$ & 4 \\
    14 & \hex{7e686868} & $x_{6} = x_{4} \land x_{5}, x_{7} = x_{2} \oplus x_{6}, x_{8} = x_{1} \oplus x_{6}, x_{9} = x_{7} \land x_{8}, x_{10} = x_{1} \land x_{3}, x_{11} = x_{3} \oplus x_{9}, x_{12} = x_{7} \oplus x_{10}, x_{13} = x_{11} \land x_{12}, x_{14} = x_{6} \oplus x_{13}$ & 4 \\
    15 & \hex{2208a208} & $x_{6} = x_{1} \land x_{2}, x_{7} = x_{3} \land x_{6}, x_{8} = x_{5} \land x_{7}, x_{9} = x_{4} \oplus x_{7}, x_{10} = x_{6} \oplus x_{9}, x_{11} = x_{1} \oplus x_{8}, x_{12} = x_{10} \land x_{11}$ & 4 \\
    16 & \hex{08888888} & $x_{6} = x_{1} \land x_{2}, x_{7} = x_{3} \land x_{6}, x_{8} = x_{4} \land x_{7}, x_{9} = x_{5} \land x_{8}, x_{10} = x_{6} \oplus x_{9}$ & 4 \\
    17 & \hex{88888888} & $x_{3} = x_{1} \land x_{2}$ & 1 \\
    18 & \hex{ea404040} & $x_{6} = x_{2} \land x_{3}, x_{7} = x_{4} \land x_{5}, x_{8} = x_{6} \oplus x_{7}, x_{9} = x_{1} \land x_{8}, x_{10} = x_{6} \oplus x_{9}$ & 3 \\
    19 & \hex{2a802a80} & $x_{5} = x_{2} \land x_{3}, x_{6} = x_{4} \oplus x_{5}, x_{7} = x_{1} \land x_{6}$ & 2 \\
    20 & \hex{73d28c88} & $x_{6} = x_{1} \land x_{2}, x_{7} = x_{2} \oplus x_{5}, x_{8} = x_{6} \oplus x_{7}, x_{9} = x_{4} \land x_{8}, x_{10} = x_{5} \oplus x_{9}, x_{11} = x_{1} \oplus x_{3}, x_{12} = x_{10} \land x_{11}, x_{13} = x_{6} \oplus x_{9}, x_{14} = x_{12} \oplus x_{13}$ & 3 \\
    21 & \hex{ea808080} & $x_{6} = x_{2} \land x_{3}, x_{7} = x_{1} \oplus x_{6}, x_{8} = x_{4} \land x_{5}, x_{9} = x_{1} \oplus x_{8}, x_{10} = x_{7} \land x_{9}, x_{11} = x_{1} \oplus x_{10}$ & 3 \\
    22 & \hex{a28280a0} & $x_{6} = x_{3} \land x_{4}, x_{7} = x_{5} \oplus x_{6}, x_{8} = x_{1} \oplus x_{2}, x_{9} = x_{7} \land x_{8}, x_{10} = x_{3} \oplus x_{9}, x_{11} = x_{1} \land x_{10}$ & 3 \\
    23 & \hex{13284c88} & $x_{6} = x_{1} \land x_{3}, x_{7} = x_{1} \oplus x_{6}, x_{8} = x_{2} \land x_{7}, x_{9} = x_{4} \oplus x_{6}, x_{10} = x_{5} \oplus x_{8}, x_{11} = x_{2} \oplus x_{10}, x_{12} = x_{9} \land x_{11}, x_{13} = x_{8} \oplus x_{12}$ & 3 \\
    24 & \hex{a2220888} & $x_{6} = x_{3} \land x_{4}, x_{7} = x_{2} \land x_{6}, x_{8} = x_{2} \oplus x_{7}, x_{9} = x_{5} \oplus x_{8}, x_{10} = x_{1} \land x_{9}$ & 3 \\
    25 & \hex{aae6da80} & $x_{6} = x_{2} \land x_{3}, x_{7} = x_{5} \oplus x_{6}, x_{8} = x_{1} \oplus x_{2}, x_{9} = x_{7} \land x_{8}, x_{10} = x_{6} \oplus x_{9}, x_{11} = x_{3} \oplus x_{10}, x_{12} = x_{5} \land x_{11}, x_{13} = x_{3} \oplus x_{12}, x_{14} = x_{1} \oplus x_{13}, x_{15} = x_{4} \land x_{14}, x_{16} = x_{9} \oplus x_{15}, x_{17} = x_{6} \oplus x_{16}$ & 4 \\
    26 & \hex{58d87888} & $x_{6} = x_{1} \land x_{2}, x_{7} = x_{1} \oplus x_{4}, x_{8} = x_{3} \land x_{7}, x_{9} = x_{3} \oplus x_{8}, x_{10} = x_{5} \land x_{9}, x_{11} = x_{3} \oplus x_{6}, x_{12} = x_{1} \oplus x_{10}, x_{13} = x_{11} \land x_{12}, x_{14} = x_{8} \oplus x_{13}$ & 4 \\
    27 & \hex{8c88ac28} & $x_{6} = x_{3} \land x_{5}, x_{7} = x_{2} \oplus x_{6}, x_{8} = x_{1} \land x_{7}, x_{9} = x_{2} \land x_{4}, x_{10} = x_{3} \oplus x_{8}, x_{11} = x_{1} \oplus x_{9}, x_{12} = x_{10} \land x_{11}, x_{13} = x_{9} \oplus x_{12}$ & 4 \\
    28 & \hex{8880f880} & $x_{6} = x_{1} \land x_{2}, x_{7} = x_{3} \oplus x_{6}, x_{8} = x_{4} \land x_{7}, x_{9} = x_{5} \land x_{8}, x_{10} = x_{6} \oplus x_{9}, x_{11} = x_{3} \land x_{10}, x_{12} = x_{8} \oplus x_{11}$ & 4 \\
    29 & \hex{9ee8e888} & $x_{6} = x_{1} \land x_{2}, x_{7} = x_{3} \oplus x_{6}, x_{8} = x_{4} \oplus x_{7}, x_{9} = x_{5} \land x_{8}, x_{10} = x_{3} \land x_{4}, x_{11} = x_{9} \oplus x_{10}, x_{12} = x_{1} \oplus x_{5}, x_{13} = x_{2} \oplus x_{12}, x_{14} = x_{9} \oplus x_{13}, x_{15} = x_{11} \land x_{14}, x_{16} = x_{6} \oplus x_{15}$ & 4 \\
    30 & \hex{4268c268} & $x_{6} = x_{2} \land x_{3}, x_{7} = x_{5} \land x_{6}, x_{8} = x_{2} \oplus x_{4}, x_{9} = x_{6} \oplus x_{8}, x_{10} = x_{1} \oplus x_{7}, x_{11} = x_{9} \land x_{10}, x_{12} = x_{3} \oplus x_{11}, x_{13} = x_{1} \land x_{12}, x_{14} = x_{6} \oplus x_{13}$ & 4 \\
    31 & \hex{16704c80} & $x_{6} = x_{1} \land x_{5}, x_{7} = x_{2} \oplus x_{6}, x_{8} = x_{4} \land x_{7}, x_{9} = x_{1} \land x_{2}, x_{10} = x_{5} \oplus x_{9}, x_{11} = x_{3} \land x_{10}, x_{12} = x_{8} \oplus x_{11}$ & 4 \\
    32 & \hex{78888888} & $x_{6} = x_{3} \land x_{4}, x_{7} = x_{5} \land x_{6}, x_{8} = x_{1} \land x_{2}, x_{9} = x_{7} \oplus x_{8}$ & 3 \\
    33 & \hex{4966bac0} & $x_{6} = x_{1} \land x_{4}, x_{7} = x_{2} \land x_{3}, x_{8} = x_{5} \oplus x_{6}, x_{9} = x_{1} \oplus x_{7}, x_{10} = x_{8} \land x_{9}, x_{11} = x_{3} \oplus x_{5}, x_{12} = x_{4} \oplus x_{6}, x_{13} = x_{2} \oplus x_{12}, x_{14} = x_{11} \land x_{13}, x_{15} = x_{10} \oplus x_{14}$ & 4 \\
    34 & \hex{372840a0} & $x_{6} = x_{1} \land x_{2}, x_{7} = x_{1} \oplus x_{6}, x_{8} = x_{3} \land x_{7}, x_{9} = x_{2} \land x_{3}, x_{10} = x_{8} \oplus x_{9}, x_{11} = x_{5} \oplus x_{10}, x_{12} = x_{4} \oplus x_{6}, x_{13} = x_{11} \land x_{12}, x_{14} = x_{8} \oplus x_{13}$ & 4 \\
    35 & \hex{5208d288} & $x_{6} = x_{1} \land x_{2}, x_{7} = x_{5} \land x_{6}, x_{8} = x_{4} \oplus x_{7}, x_{9} = x_{1} \oplus x_{3}, x_{10} = x_{8} \land x_{9}, x_{11} = x_{6} \oplus x_{7}, x_{12} = x_{10} \oplus x_{11}$ & 3 \\
    36 & \hex{7ca00428} & $x_{6} = x_{1} \oplus x_{4}, x_{7} = x_{2} \land x_{6}, x_{8} = x_{5} \land x_{7}, x_{9} = x_{1} \oplus x_{5}, x_{10} = x_{7} \oplus x_{9}, x_{11} = x_{4} \land x_{10}, x_{12} = x_{2} \oplus x_{3}, x_{13} = x_{1} \oplus x_{11}, x_{14} = x_{12} \land x_{13}, x_{15} = x_{8} \oplus x_{14}$ & 4 \\
    37 & \hex{f8880888} & $x_{6} = x_{1} \land x_{2}, x_{7} = x_{5} \oplus x_{6}, x_{8} = x_{3} \land x_{7}, x_{9} = x_{4} \land x_{8}, x_{10} = x_{6} \oplus x_{9}$ & 3 \\
    38 & \hex{2ec0ae40} & $x_{6} = x_{3} \oplus x_{4}, x_{7} = x_{2} \land x_{6}, x_{8} = x_{4} \oplus x_{7}, x_{9} = x_{1} \land x_{8}, x_{10} = x_{5} \land x_{9}, x_{11} = x_{7} \oplus x_{10}, x_{12} = x_{2} \land x_{11}, x_{13} = x_{9} \oplus x_{12}$ & 4 \\
    39 & \hex{f888f888} & $x_{5} = x_{1} \land x_{2}, x_{6} = x_{3} \land x_{4}, x_{7} = x_{5} \land x_{6}, x_{8} = x_{5} \oplus x_{7}, x_{9} = x_{6} \oplus x_{8}$ & 3 \\
    40 & \hex{58362ec0} & $x_{6} = x_{3} \land x_{5}, x_{7} = x_{4} \oplus x_{5}, x_{8} = x_{6} \oplus x_{7}, x_{9} = x_{1} \oplus x_{2}, x_{10} = x_{8} \land x_{9}, x_{11} = x_{4} \oplus x_{10}, x_{12} = x_{1} \land x_{11}, x_{13} = x_{3} \oplus x_{12}, x_{14} = x_{2} \oplus x_{6}, x_{15} = x_{13} \land x_{14}, x_{16} = x_{10} \oplus x_{15}$ & 4 \\
    41 & \hex{0eb8f6c0} & $x_{6} = x_{3} \land x_{5}, x_{7} = x_{1} \land x_{4}, x_{8} = x_{5} \oplus x_{7}, x_{9} = x_{1} \oplus x_{6}, x_{10} = x_{8} \land x_{9}, x_{11} = x_{2} \oplus x_{7}, x_{12} = x_{4} \oplus x_{11}, x_{13} = x_{2} \oplus x_{10}, x_{14} = x_{3} \oplus x_{6}, x_{15} = x_{13} \oplus x_{14}, x_{16} = x_{12} \land x_{15}, x_{17} = x_{6} \oplus x_{16}, x_{18} = x_{11} \oplus x_{17}$ & 4 \\
    42 & \hex{567cea40} & $x_{6} = x_{1} \land x_{4}, x_{7} = x_{5} \land x_{6}, x_{8} = x_{2} \oplus x_{7}, x_{9} = x_{1} \oplus x_{8}, x_{10} = x_{3} \land x_{9}, x_{11} = x_{5} \oplus x_{10}, x_{12} = x_{2} \oplus x_{3}, x_{13} = x_{11} \land x_{12}, x_{14} = x_{6} \oplus x_{10}, x_{15} = x_{13} \oplus x_{14}$ & 4 \\
    43 & \hex{f8887888} & $x_{6} = x_{1} \land x_{2}, x_{7} = x_{3} \land x_{4}, x_{8} = x_{6} \land x_{7}, x_{9} = x_{5} \land x_{8}, x_{10} = x_{6} \oplus x_{9}, x_{11} = x_{7} \oplus x_{10}$ & 4 \\
    44 & \hex{78887888} & $x_{5} = x_{1} \land x_{2}, x_{6} = x_{3} \land x_{4}, x_{7} = x_{5} \oplus x_{6}$ & 2 \\
    45 & \hex{e72890a0} & $x_{6} = x_{2} \land x_{5}, x_{7} = x_{3} \oplus x_{6}, x_{8} = x_{1} \land x_{7}, x_{9} = x_{2} \oplus x_{4}, x_{10} = x_{3} \land x_{9}, x_{11} = x_{5} \oplus x_{10}, x_{12} = x_{4} \land x_{11}, x_{13} = x_{8} \oplus x_{12}$ & 4 \\
    46 & \hex{268cea40} & $x_{6} = x_{2} \land x_{3}, x_{7} = x_{4} \oplus x_{6}, x_{8} = x_{1} \land x_{7}, x_{9} = x_{2} \land x_{5}, x_{10} = x_{6} \oplus x_{8}, x_{11} = x_{9} \oplus x_{10}$ & 3 \\
    47 & \hex{6248eac0} & $x_{6} = x_{1} \land x_{4}, x_{7} = x_{3} \land x_{6}, x_{8} = x_{5} \oplus x_{7}, x_{9} = x_{1} \land x_{8}, x_{10} = x_{3} \oplus x_{9}, x_{11} = x_{2} \land x_{10}, x_{12} = x_{6} \oplus x_{11}$ & 4 \\
    \bottomrule
  \end{tabularx}
\end{table*}

\end{document}